\documentclass{osa-article}

\journal{oe}
\articletype{Research Article}
\usepackage{bm}
\usepackage{lineno}
\def\ii{{\mathrm{i}}}

\usepackage{gensymb}
\usepackage{here}
\usepackage{svg}

\begin{document}

\title{Isolation of phase edges using off-axis q-plate filters}

\author{Jigme Zangpo\authormark{*} and Hirokazu Kobayashi.}

\address{Graduate School of Engineering, Kochi University of Technology, 185 Miyanokuchi, Tosayamada, Kami City, Kochi 782-8502, Japan}

\email{\authormark{*}jigmezangpo11@gmail.com} %% email address is required



\begin{abstract}
Edge-enhanced microscopes with q-plate have been attracted more attention to enhance the edges of phase-amplitude objects in biological sample due to their capability for all-directional edge enhancement, while differential interference-contrast microscopy enhances edges in only one-direction. However, the edge-enhanced microscopes cannot distinguish the edges of phase and amplitude objects, as both edges are equally enhanced. This study introduces a novel method for isolating the edge of a phase object from an amplitude object using an off-axis q-plate filter in a $4f$ system. Herein, we combined off-axis q-plates with four different displacements to isolate the phase object edge from the amplitude object. To demonstrate the proposed method, we conducted experiments using two distinct samples. The first sample comprised a phase test target surrounded by an aperture, and the second sample involved an overlap between the phase test target and a white hair with non-zero transmittance. In the samples, the isolated phase object edge is in good agreement with the theoretical expectations, and the amplitude object edge was reduced by approximately 93\%. The proposed method is a novel and effective approach for isolating the edge of a phase object from an amplitude object and can be useful in various biological imaging applications.
%The first sample consisted of a phase test target mixed with an aperture, while the second sample involved a phase test target mixed with a human white hair. The phase test target served as the phase object, while the aperture and white hair, with zero and non-zero transmittance, respectively, served as the amplitude objects. In the first sample, there was no overlap between the phase object and amplitude object, while the second sample involved an overlap between the two. In both samples, the retrieved phase object edge reduced the amplitude object edge by approximately 93\% compared to the on-axis q-plate image. The experimental results agree well with the theoretical expectations, and the numerical simulation results.
\end{abstract}

%%%%%%%%%%%%%%%%%%%%%%%%%%  body  %%%%%%%%%%%%%%%%%%%%%%%%%%
\section{Introduction}
In recent years, the biological and medical research field has witnessed the development of increasingly sophisticated optical microscopes. The most traditional optical microscope is the bright-field microscope, which can observe opaque amplitude objects (AOs) but provide low-contrast images of transparent phase objects (POs)\cite{khandpur2020brightfield,stagaman1988bright}. In the 1930s, the first phase-contrast microscope invented by Dutch physicist Frits Zernike could visualize the POs that were not detected by the traditional bright-field microscope\cite{burch1942phase,frohlich2008phase,kohler1947phase,zernike1942phase}. In the 1950s, Georges Nomarski, a Polish physicist and optics theorist, developed a differential interference-contrast microscope, which highlights PO structural details and contributes to edge enhancement\cite{lang1982nomarski,frohlich2008phase}. Edge enhancement plays an important role in optical information processing that highlights the contours of an object and is widely used in image processing \cite{offaxis1, imageprocessing2}, microscopy and biological imaging \cite{biological1, offaxis2, biological3}, fingerprint recognition \cite{fingerprint1}, the medical field \cite{medical}, and astronomy \cite{astonomy1,astonomy2}. Despite the recognition of differential interference-contrast microscope as well-established phase imaging techniques, it enhances edges of the objects in only one-direction\cite{DIC1, DIC2, DIC3}. 

Ritsch-Marte \textit{et al.} used a state-of-the-art vortex filter placed on the Fourier plane of a $4f$ system, called edge-enhanced microscopy, to overcome the limitations of differential interference contrast microscopy and to observe all-directional edges of POs and AOs\cite{spiralPF2,spiralPF4,spiralPF7,spiralPF8}. Vortex filters are classified as scalar vortex filtering (SVF) and vectorial vortex filtering (VVF). SVF is a polarization-independent vortex filter that generates a spatial phase distribution of $e^{i\phi}$, where $\phi$ denotes the azimuthal angle, while VVF is a polarization-dependent vortex filter that generates with azimuthal phase with opposite sign depending on the helicity of the input circular polarization. When a spiral phase filter is generated by a spatial light modulator \cite{offaxis3, SLM2, SLM3, spiralPF8} or a vortex phase plate \cite{RCP1, RCP2} placed in a $4f$ system, it exhibits SVF, whereas a spatially-variable half-waveplate, such as a q-plate or an s-wave plate, placed in a $4f$ system exhibits VVF\cite{qplate1, qplate2, qplate3,swaveplate1, swaveplate2, swaveplate3}. A biological specimens are not only pure POs but also phase-amplitude objects (PAOs) containing PO and AO, and the enhanced edge of a PAO reveals the indistinguishable edges of a PO and AO\cite{zangpo2023edge}. Thus, for improving the visualization of the PAO, it has attracted much attention in biological imaging to isolate PO edges from an AO. Although SVF and VVF can enhance PAO edges, no approach has been proposed to isolate the edges of the PO from the AO so far.  
%The Transport of Intensity Equation (TIE), a noninterferometric technique that requires several observations along the propagation direction, was used for phase retrieval\cite{TIE1defocus,TIE2tilted,TIE4Hilbert,TIE5,TIE3singleshot}. While TIE-based phase retrieval techniques have been used to retrieve phases from PO, little quantitative research on PO retrieval from PAO has been presented to date. 

Herein, we propose and demonstrate the isolation of PO edges from an AO using off-axis q-plates. Several research papers have reported that an off-axis vortex filter can be used to enhance edges anisotropically via SVF\cite{offaxis1,offaxis2,offaxis3,offaxis4,offaxis5PC-SPF,spiralPF2}. To the best of our knowledge, however, no researcher has used an off-axis vortex filter to isolate PO edges from an AO.
In our previous research, we observed biological sample, an onion cell using both SVF and VVF. It shows that VVF has an advantage over SVF as there was no interference between PO and AO in the PAO. Despite VVF enhance edges well, the edges of PO and AO are indistinguishable because they are equally enhanced. According to \cite{zangpo2022isolation,zangpo2023edge}, the theoretical analysis and numerical simulation indicate that only VVF can isolate PO edges from an AO by combining the images obtained using four tilted incident lights. In this study, to simplify the experimental setup, off-axis q-plate filter instead of tilting the incident light is used in the proof-of-principle experiment. We combine an off-axis q-plate filter with four different displacements to isolate PO edges from an AO. Two PAO samples are used to demonstrate the proposed method of isolating PO edges from an AO. The first sample comprises a phase test target surrounded by an aperture, whereas the second sample involves an overlap between the phase test target and a white hair with non-zero transmittance. In the samples, the isolated PO edge agrees well with the theoretical expectations, and the AO edge is reduced by approximately 93\%. Moreover, the discussion section presents numerical simulations that optimize displacement of off-axis q-plate and recover phase edges overlapped with the AO using deconvolution method. 

In Section 2 we describe the theoretical analysis of isolating PO edges from an AO. In Sections 3 and 4, we present the experimental setup and results, respectively. Section 5 includes the discussion part and Section 6 concludes the study.

\section{Theoretical analysis}

\begin{figure}[htp!]
\centering
\includegraphics[width=\linewidth]{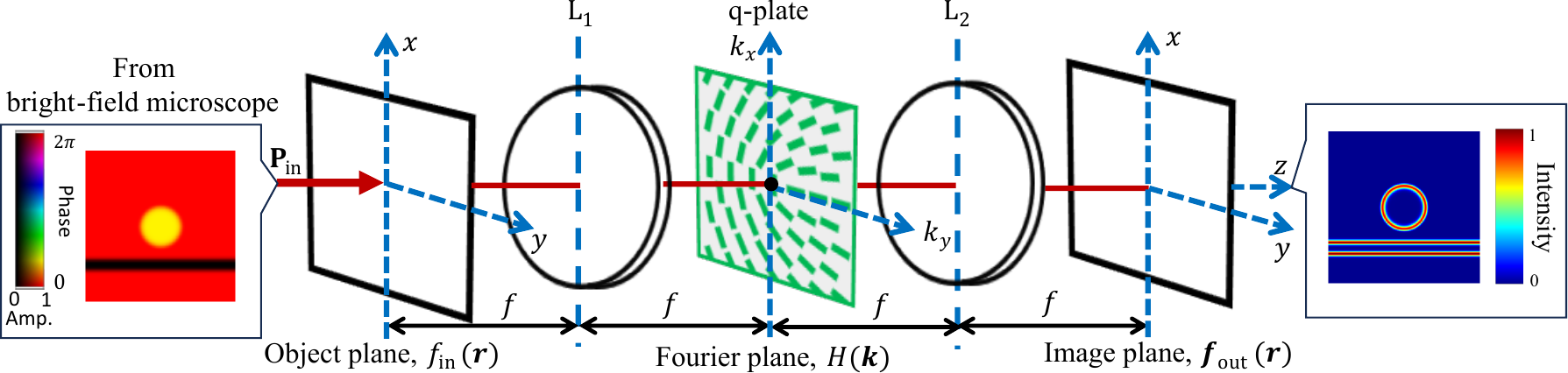}
\caption{$4f$ imaging system with an on-axis q-plate.}
\label{fig1}
\end{figure}

First, we explain that the conventional $4f$ system with an on-axis q-plate filter on the Fourier plane can implement all-directional edge enhancement. The setup in Fig. \ref{fig1} is placed just after the bright-field microscope, where the magnified object located on the object plane undergoes edge enhancement by the $4f$ system with q-plate, as shown on the image plane.
%In off-axis q-plate, the green dot represents the center of the propagation axis, and the dot with the purple color is the center of the q-plate, which displace from the center of propagation. 
The input object $f_{\text{in}} (\bm{r})$, with $\bm{r}$ being two-dimensional position vector on transverse coordinate $(x,y)$, is assumed to be illuminated by the incident horizontally-polarized beam with Jones's vector $\bm{P}_\text{in}=\big(\begin{smallmatrix}
  1\\0\end{smallmatrix}\big)$. The polarized light through the object undergoes Fourier transform by the lens $\text{L}_\text{1}$ resulting in $F_{\text{in}}(\bm{k})\bm{P}_\text{in}$ on the Fourier plane, where $F_\text{in}(\bm{k})$ denotes the Fourier transform of the object with $\bm{k}=(k_x,k_y)$ being two-dimensional transverse wavenumber vector. The on-axis q-plate filter is placed on the Fourier plane at the front focal plane of lens $\text{L}_2$.

The transmission function of the on-axis q-plate\cite{q-platetheory1,q-platetheory2,q-platetheory3} with retardance $\pi$ radians in polar coordinate is defined by the Jones matrix, as shown below:

\begin{equation}
    H_q(\bm{r})=\ii\begin{pmatrix}
    \text{cos}(2q\theta)& \text{sin}(2q\theta)\\
    \text{sin}(2q\theta)& -\text{cos}(2q\theta)
    \end{pmatrix},
    \label{Eq1}
\end{equation}
where $\theta=\text{tan}^{-1}(y/x)$ and the q-plate with topological charge $q$ can generate a $\pm2q$\ charge optical vortex depending on the helicity of the input circular polarization. The transmission function of the q-plate in Cartesian coordinate derived from Eq. (\ref{Eq1}) by setting $q=1/2$ is expressed as follows,
\begin{equation}
    H_{1/2}(\bm{r}) =\frac{\ii}{r} 
    \begin{pmatrix}
    x& y\\
    y& -x
    \end{pmatrix},
    \label{Eq2}
\end{equation}
where ${r}=\sqrt{\smash{x}^2+\smash{y}^2}$. 

%Considering illumination source, $\bm{P}_\text{in}=\big(\begin{smallmatrix}1\\0\end{smallmatrix}\big)$ 
%The incident beam $\bm{P}_\text{in}$ illuminating the object $f_{\text{in}} (\bm{r})$, and 

%The illuminated object undergoes a Fourier transform at $\text{L}_1$. The object spectrum of the Fourier transform defined as ${F}_\text{in}(\bm{k})\bm{P}_\text{in}$, where ${F}_\text{in}(\bm{k})$ denotes the Fourier transform of the input object. 

When the object spectrum reach the filter plane, it multiplies with the transmission function of the filter $H_{1/2}(\bm{k})$ and the modulated spectrum is expressed as ${F}_\text{in}(\bm{k})H_{1/2}(\bm{k})\bm{P}_\text{in}$. Then, the modulated spectrum undergoes an inverse Fourier transform $\mathcal{F}^{-1}$ by the lens $\text{L}_2$ and complex amplitude $\bm{f_{\text{out}}} (\bm{r})$ with polarization state on the image plane located at the rear focal plane of lens $\text{L}_2$ is $\bm{f_{\text{out}}} (\bm{r})=\mathcal{F}^{-1}\{F_\text{in}(\bm{k})H_{1/2}(\bm{k})\bm{P}_\text{in}\}$. By simplifying $\bm{f_{\text{out}}} (\bm{r})$ based on differential property of Fourier transform, i.e., ${k}_x {F}_\text{in}({k}_x)\leftrightarrow-\ii\partial_x f_\text{in}(x)$, the output of the $4f$ system with an on-axis q-plate can be obtained as
\begin{equation}
  \bm{f}_{\text{out}} (\bm{r})
    = \frac{1}{r}\ast \bm{\nabla} f_{\text{in}} (\bm{r}),  
  \label{Eq2.1}
\end{equation}
where $\bm{\nabla}=(\frac{\partial}{\partial x},\frac{\partial}{\partial y})$ and $\ast$ denotes convolution. Although $\bm{\nabla} f_{\text{in}}(\bm{r})$ generates the edges of an object, the convolution term $1/r$ does not directly contribute to the detection of the edges of the object; however, causes edge broadening. On the CMOS camera, the obtained output intensity $I(\bm{r})$ is
\begin{equation}
  I(\bm{r})
  =|\bm{f}_\text{out} (\bm{r})|^2
    = \left|\frac{1}{r}\ast \bm{\nabla} f_{\text{in}} (\bm{r})\right|^2.  
  \label{abs}
\end{equation}

When considering the output with the off-axis q-plate, the center of the q-plate needs to be shifted from the origin of the filter plane to a new point, i.e., $\bm{k}_0=(k_{x0},k_{y0})$; then, the new transmission function of the q-plate becomes $H(\bm{k}-\bm{k}_0)$. According to Refs. \cite{zangpo2022isolation, zangpo2023edge, qplate2}, to isolate PO edges from an AO, a q-plate filter needs to be generated via VVF and incident illumination should be linearly polarized. Owing to the displacement of the q-plate, it generates an extra phase and the output on the image plane is given by

\begin{equation}
  \bm{f}_\text{out} (\bm{r}, \bm{k}_0)
  =\mathcal{F}^{-1}\{F_\text{in}(\bm{k})H_{1/2}(\bm{k}-\bm{k}_0)\bm{P}_\text{in}\}
  =e^{\ii \bm{k}_0\cdot \bm{r}} \mathcal{F}^{-1}\{F_\text{in}(\bm{k}+\bm{k}_0)H_{1/2}(\bm{k})\bm{P}_\text{in}\}.
  \label{Eq4}
\end{equation}

Applying differential and frequency shifting property of Fourier transform, i.e., ${k}_x F_\text{in}({k}_x+k_{x0})\leftrightarrow-\ii \partial_x f_\text{in}(x)e^{-\ii k_{x0}x}$, we obtain the output as follows:

\begin{equation}
  \bm{f}_\text{out} (\bm{r},\bm{k}_0)
  =e^{\ii \bm{k}_0\cdot \bm{r}} \left\{\frac{1}{r} \ast \bm{\nabla} f_{\text{in}} (\bm{r}) e^{-\ii \bm{k}_0\cdot \bm{r}} \right\},
  \label{Eq5}
\end{equation}
where the extra phase ($\bm{k}_0\cdot \bm{r}$) is contributed by the off-axis q-plate. 

Given a complex specimen $f_{\text{in}} (\bm{r}) = A(\bm{r})\text{Exp}[\ii B(\bm{r})]$ with an amplitude function $A(\bm{r})$ and a phase function $B(\bm{r})$, the edge enhancement of the complex specimen $f_{\text{in}} (\bm{r})$ can be obtained on the CMOS camera by calculating absolute squared value of Eq. (\ref{Eq5}) as follows:

\begin{equation}
  I(\bm{r},\bm{k}_0)=|f_\text{out} (\bm{r},\bm{k}_0)|^2 \xrightarrow{\text{ideal edge}}
    \left|\bm{\nabla}A(\bm{r})\right|^2
    +A(\bm{r})^2\left|\bm{\nabla}B(\bm{r})-\bm{k}_0\right|^2,
    \label{Eq51}
\end{equation} 
where $\xrightarrow{\text{ideal edge}}$ indicates that the $1/r$ convolution is omitted, resulting in ideal edges without broadening effect of the $1/r$ convolution. The ideal approximation in Eq. (\ref{Eq51}) will sharpen the edges of the object because of no convolution term $1/r$, however, on the CMOS camera in the experiment results and also in our simulation results, the image includes the convolution term $1/r$. The effect of the convolution term $1/r$ will be discussed in our experimental and simulation results. The first and second terms on the right side of Eq. (\ref{Eq51}) represent the AO and PO edges, respectively. As only the second term comprises the contribution of the off-axis q-plate, the AO edges in the first term can be eliminated by combining the outputs with four different off-axis q-plates and calculating the following equation:

\begin{equation}
\begin{split}
    J(\bm{r},{k}_0)&\equiv\left|\left\{
        I(\bm{r},k_0\bm{e}_x)-I(\bm{r},-k_0\bm{e}_x,)
        \right\}\right|^2
        +\left|\left\{
        I(\bm{r},k_0\bm{e}_y)-I(\bm{r},-k_0\bm{e}_y)
        \right\}\right|^2\\
    &\xrightarrow{\text{ideal edge}} \left|4 k_0 A(\bm{r})^2\bm{\nabla}B(\bm{r})\right|^2,
    \end{split}
    \label{Eq6}
\end{equation}
where $k_0$ denotes the displacement of the q-plate in Fourier plane, and $\bm{e}_x$ and $\bm{e}_y$ denote unit vectors along $x$ and $y$ axes, respectively. Equation (\ref{Eq6}) represents the approximately-isolated PO edges from an AO using the off-axis q-plates, where the convolution term with $1/r$ was omitted. The multiplication factor in front of the isolated PO edges, i.e., $16 k_0^2 A(\bm{r})^4$, may obstruct in isolating the PO edge depending on the transmittance of the AO if the PO is overlapped with the AO. Here we note that  the positive and the negative displacements in the horizontal or vertical direction with the amount of $k_0$ have to be symmetric about the optical axis determined in the $4f$ system to accurately isolate the PO edge. If the q-plate is asymmetrically displaced, the isolated PO edge becomes non-uniform because of the asymmetric extra phases induced on the PO, leading to varying intensities on the isolated PO edges. 

%The q-plate should be symmetric about the center to isolate the PO edges. If the q-plate is not symmetric about the center, then the amount of shift $k_0$ will be different between the positive and negative shifts in the horizontal (or vertical) direction, resulting in non-uniform intensity on the isolated PO edge. This occurs because the different shifts $k_0$ result in different extra phases induced on the PO, leading to varying intensities on the isolated PO edges. Furthermore, the reduction of AO edges cannot be achieved due to the different intensities of AO edges resulting from different shift $k_0$ of the filter when the q-plate is not symmetric about the centre. 

\section{Experimental setup}
\begin{figure}[htp!]
\centering
\includegraphics[width=0.9\linewidth]{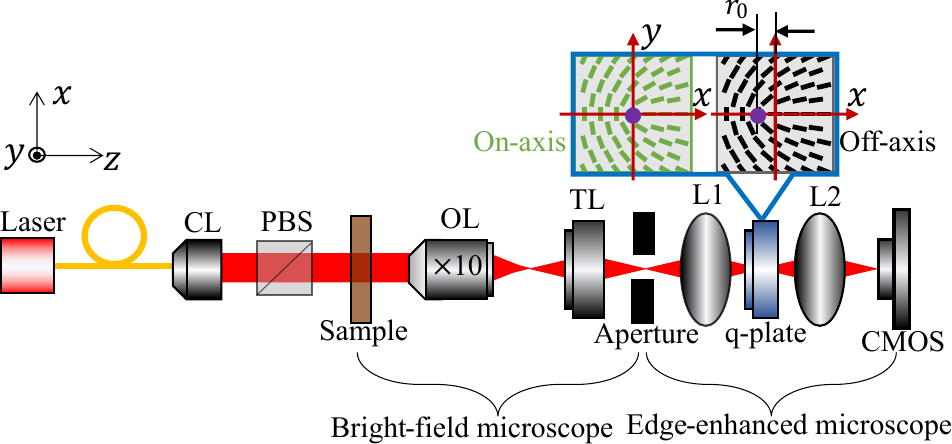}
\caption{Experimental setup. CL denotes the collimation lens; PBS denotes the polarizing beam splitter; OL denotes the objective lens; TL denotes the tube lens; L1 and L2 denote the Fourier lenses; CMOS denotes the complementary metal-oxide-semiconductor camera. $f_{\text{CL},\text{OB},\text{TL},1,2}=50,18,180,300,$ and $250$ mm, the subscripts denote the names of the lenses.}
\label{fig2}
\end{figure}

The experimental setup for isolating PO edges from an AO via edge-enhanced microscopy is illustrated in Fig. \ref{fig2}, consisting of a bright-field microscope and a $4f$ imaging system with a q-plate to enhance the edge of an object. In Refs. \cite{qplate2,zangpo2022isolation, zangpo2023edge}, on-axis VVF was considered as the vortex filter, whereas herein, we employed off-axis VVF. The light field of a laser (Thorlabs, HLS635) with a wavelength of 635 nm  was collimated using a collimation lens (Thorlabs, AC254-050-A-ML) and converted to horizontal linear polarization using a polarizing beam splitter (Thorlabs, CM1-PBS251). The first $4f$ system represents bright-filed microscopy comprising an objective lens (focal length, $f_{\text{OB}}=18$ mm) with a magnification factor of 10 and a tube lens ($f_{\text{TL}}=180$ mm); the second $4f$ system includes a commercially available q-plate (Thorlabs, WPV10L-633) with retardance $\pi$ and topological charge $1$ located on the Fourier plane, which is between the midpoint of the rear focal length ($f_1=300$ mm) of Lens L1 and the front focal length ($f_2=250$ mm) of Lens L2. The q-plate was displaced at a distance of $r_0$ in four directions ($+x$, $-x$, $+y$, and $-y$), where $r_0=\frac{f_1}{k}k_0$ with $k$ being the wavenumber and $k_0$ being the wavenumber displacement on the Fourier plane. Figure \ref{fig2} shows one of the four off-axis q-plate displaced in the $-x$ direction, with the purple dot representing the center of the q-plate. A CMOS camera (Thorlabs, DCC1645C) was placed on the image plane, which was located at the rear focal length position of Lens L2. The camera had an imaging area of $4.6\text{ mm}\times3.7\text{ mm}$ and a pixel size of 3.6 \textmu m $\times$3.6 \textmu m.

We first observed the sample using an on-axis q-plate. Then, to isolate the PO edge, we displaced the q-plates in four different directions: two in the $x$ direction ($+x$ and $-x$) and two in the $y$ direction ($+y$ and $-y$). Hereafter, four different images recorded using the CMOS camera were combined according to Eq. (\ref{Eq6}). If the q-plate is not symmetric about the center after being displaced in one direction, then the amount of shift $r_0$ will be different between the positive and negative shifts in the $x$ or $y$ direction. Therefore, we employed an $xy$ translator with a differential drive to achieve precise q-plate positioning. The differential drive enables fine adjustments down to 0.5 \textmu m per graduation, surpassing the 10 \textmu m per graduation of a micrometer drive. The commercially available phase test target (Benchmark Technologies, Quantitative Phase Target) included USAF 1951 resolution targets with heights of 50, 100, 150, 200, 250, 300, and 350 nm with a refractive index of 1.52. Moreover, the phase test target included the letters "IES." The first PAO sample comprised the 350-nm USAF 1951 resolution target surrounded by an aperture, where the PO (phase test target) and AO (aperture) did not overlap. In contrast, the second PAO sample was created by overlapping the phase test target ("IES") with a white hair, which has non-zero transmittance.

\section{Results}
To demonstrate our method of the PO edge isolation from the AO, we present the proof-of-principle experimental results with the two different samples in separated sections. For quantitative evaluation of our method, the correlation coefficient between the isolated PO edges of the on-axis and off-axis q-plates is determined. Further, we calculate reduction ratio of the AO edge in the isolated PO edge.

\subsection{Phase test target surrounded by a circular aperture}
\label{POaperture}

\begin{figure}[htp!]
\centering
\includegraphics[width=\linewidth]{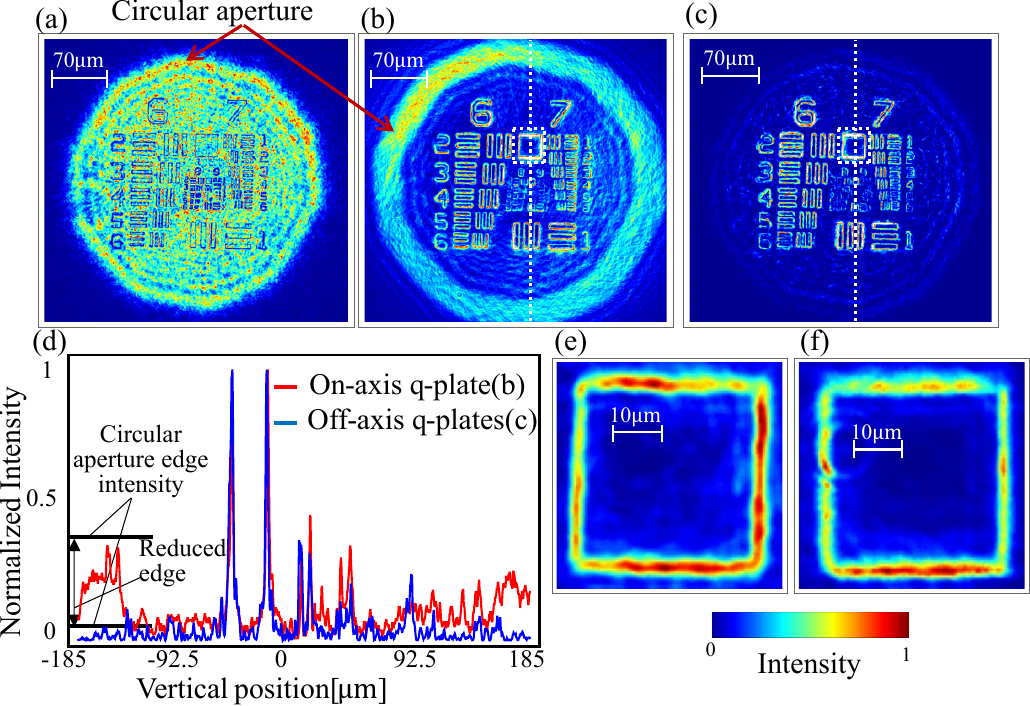}
\caption{(a) PAO without the q-plate, (b) PAO with the on-axis q-plate, (c) PAO with the off-axis q-plate, (d) vertical cross-sectional distribution of the on-axis and off-axis images, and (e) and (f) show the extracted square POs from the (b) and (c) images, respectively.}
\label{fig3}
\end{figure}

The first PAO is the 350-nm USAF 1951 resolution phase target surrounded by an circular aperture with the radius of 300 \textmu m, which acts as an AO. Figure \ref{fig3} (a) shows the PAO without a q-plate; we can clearly see the aperture; however the PO is not clearly visible. To enhance the edge of the sample, the on-axis q-plate is used, and the experimental result is shown in Fig. \ref{fig3} (b). Here, the PO and AO edges of the PAO have been enhanced, which was not possible without the q-plate, as shown in Fig. \ref{fig3} (a). To isolate the PO edge from an AO edge in Fig. \ref{fig3} (b), we displaced the q-plate along the $\pm x$ and $\pm y$ directions, with 125 \textmu m, combined the resulting four images according to Eq. (\ref{Eq6}), and obtained the isolated PO edge as shown in Fig. \ref{fig3} (c). The vertical cross-sectional intensities of the on-axis (red curve) and off-axis (blue curve) images are illustrated in Fig. \ref{fig3} (d), which indicates that the circular aperture edge has been reduced on the off-axis image.

For quantitative analysis, we defined the AO edge reduction in the off-axis q-plate image as $R=\frac{P_{\text{on}}-P_{\text{off}}}{P_{\text{on}}}$, where $P_{\text{on}}$ and $P_{\text{off}}$ denote the power of the AO edge with the on-axis and with the PO edge isolation, respectively. The AO edge reduction was $R=0.97$, which indicates that there is an evident reduction in the AO edge while viewing the off-axis image. Moreover, to analyze the image quality of the isolated PO edge, we extracted the square-shaped PO edges from the on-axis result in Fig. \ref{fig3} (b) and the PO edge isolation result in Fig. \ref{fig3} (c), as shown in Figs. \ref{fig3} (e) and (f), respectively. Their correlation coefficient was 0.93, indicating a successful isolation of the PO edge in the off-axis image.

According to Eq. (\ref{Eq6}), the displacement $r_0$ of the q-plate plays a crucial role in isolating the intensity of the PO edge. As $r_0$ increases, the correlation coefficient between the on-axis and off-axis q-plate images monotonically increases because the intensity of the PO edge is proportional to $r_0$. However, when $r_0$ considerably increases such that the q-plate moves away from the beam radius, we expect the correlation coefficient to decrease. 

\begin{figure}[htp!]
\centering
\includegraphics[width=\linewidth]{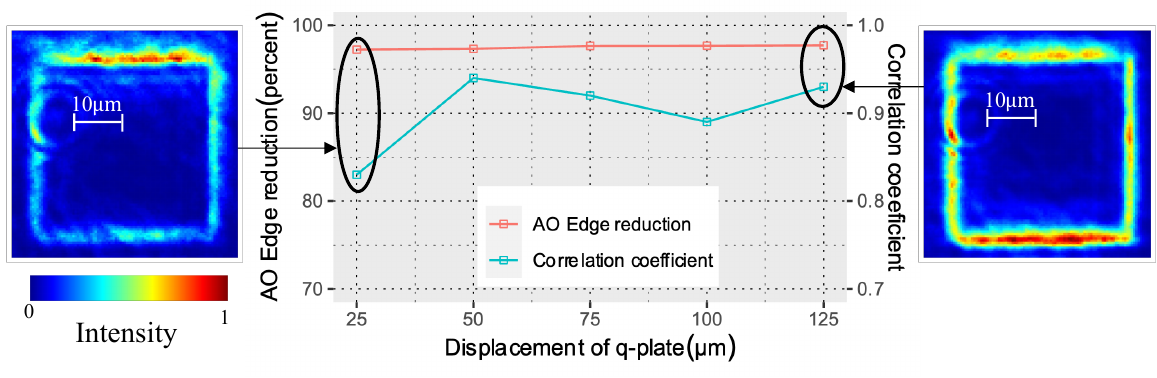}
\caption{AO edge reduction in off-axis q-plate image compared with on-axis q-plate image for different q-plate displacement and correlation coefficient between on-axis PO edge and isolated PO edge by off-axis q-plate for different q-plate displacement.}
\label{fig4}
\end{figure}

To investigate the optimized value of $r_0$, we determined the correlation coefficient between the on-axis  and off-axis q-plate images for varying $r_0$, as shown in Fig. \ref{fig4}. As $r_0$ increases, the correlation coefficient improves from 0.83 (for a 25 \textmu m shift) to 0.93 (for a 125-\textmu m shift), whereas the AO edge reduction (approximately 97\%) remains almost constant for q-plate shifts of 25-125 \textmu m. The $xy$ translator used to displace a q-plate possesses a total scale of 250 \textmu m, and a distance of $\pm$125 \textmu m can be displaced in the $\pm x$ and $\pm y$ directions, with a step of 25 \textmu m. Thus, as the full-scale is limited to 125 \textmu m, we could not experimentally determine the optimal displacement of the off-axis q-plate. We numerically simulated the optimal displacement of the off-axis q-plate under conditions similar to the experimental setup, as discussed in Section \ref{D1}.

\begin{figure}[htp!]
\centering
\includegraphics[width=\linewidth]{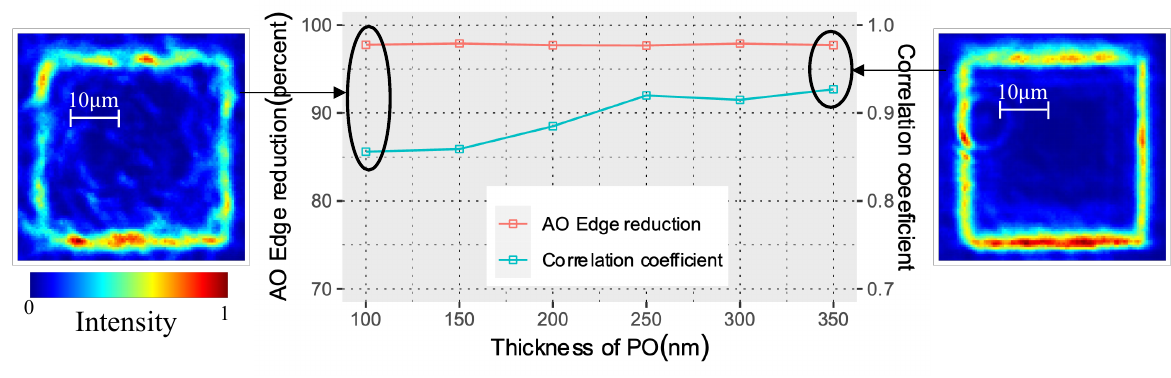}
\caption{AO edge reduction in the off-axis q-plate image compared the with on-axis q-plate image for different thicknesses of the PO and the correlation coefficient between the on-axis PO edge and PO edge isolated by the off-axis q-plate for different thicknesses of the PO.}
\label{fig41}
\end{figure}

To investigate the behavior of the isolated PO edge in the off-axis q-plate for phase gradient, we used PO thicknesses ranging from 100 to 350 nm. As shown in Fig. \ref{fig41}, the correlation coefficient improved as the thickness increased because the phase difference at 100 nm is 0.56 rad, which is lower than the phase difference at 350 nm, i.e., 1.99 rad. The lower the phase difference, the flatter the slope/gradient of the PO, and the edge intensity decreases, resulting in a low correlation coefficient.

\subsection{Phase test target overlapping with a white hair}

\begin{figure}[htp!]
\includegraphics[width=\linewidth]{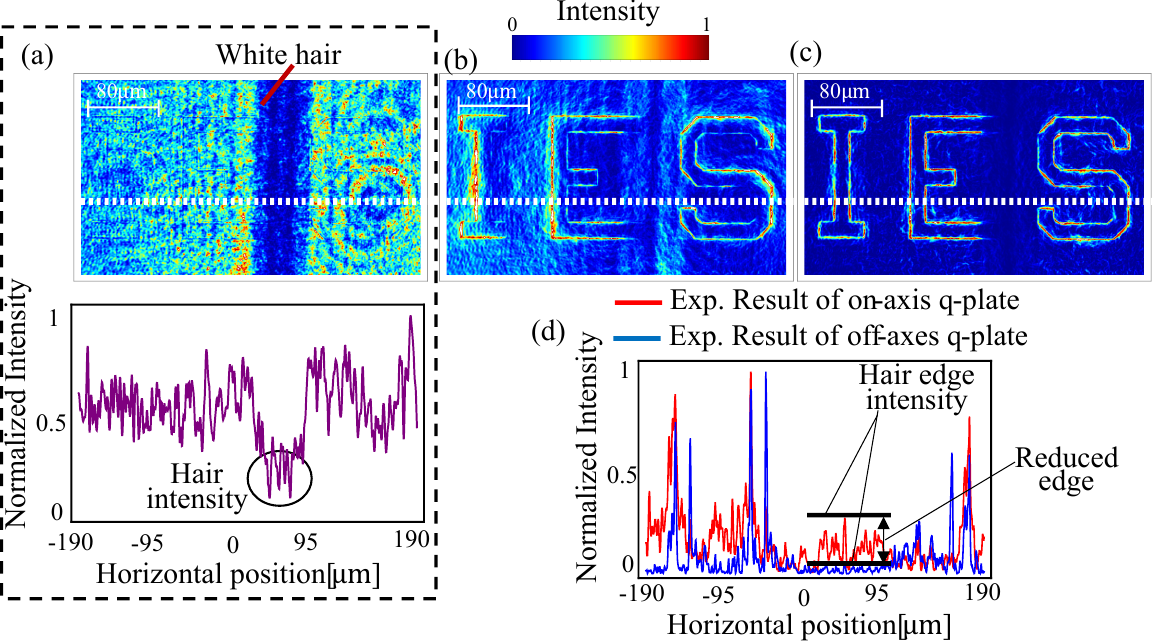}
\caption{Experimental results of the PAO (phase test target plus white hair) (a) without the q-plate and its horizontal cross-sectional intensity distribution, (b) with the on-axis q-plate, (c) off-axis q-plate with displacement of 125\textmu m, and (d) horizontal cross-sectional intensity distribution for (b) and (c).}
\label{fig5}
\end{figure}

In the first PAO, the PO did not overlap with the AO. Now, in this section, we demonstrate the experimental results of the phase test target ("IES") overlapping with a white hair, which has non-zero transmittance, unlike the circular aperture with zero transmittance. The CMOS camera was placed at imaging position on the PO, whereas the white hair was positioned out of focus at a distance of 0.8 mm from the PO.

Figure \ref{fig5} (a) shows the experimental results of the PAO without a q-plate and its horizontal cross-sectional intensity distribution. The white hair (vertical strip) is visible in Fig. \ref{fig5} (a) because it is an opaque AO, whereas the transparent PO is not visible. The transmittance of the white hair is determined as approximately 60\% from the horizontal cross-sectional intensity distribution. With the on-axis q-plate, both edges of the PO and AO can be observed, as shown in Fig. \ref{fig5} (b). However, the background intensity is nonzero because of the diffracted light from the white hair. Figure \ref{fig5} (c) shows the isolated PO edge using the off-axis q-plate based on Eq. (\ref{Eq6}). The edge of the white hair and its diffraction have reduced, as shown in Fig. \ref{fig5} (c) compared to Fig. \ref{fig5} (b), and the PO edge became clearer. Additionally, the horizontal cross-sectional intensity distributions of the on-axis (red curve) and off-axis (blue curve) q-plates shown in Fig. \ref{fig5} (d) exhibit that the edge of the white hair was reduced by approximately $93\%$ in the off-axis image. 

Despite the reduction in the white hair edge shown in Fig. \ref{fig5}(c), the PO overlapping the white hair is not clearly isolated as can be seen from the letter "E" being broken on the right-hand side.
This is because the isolated PO edge is directly proportional to the intensity of the AO, as expressed in Eq. (\ref{Eq6}). We discuss this issue using numerical simulation in Section \ref{D2}.

\section{Discussion}
\subsection{Optimization of off-axis q-plate displacement}
\label{D1}
\begin{figure}[htp!]
\centering
\includegraphics[width=\linewidth]{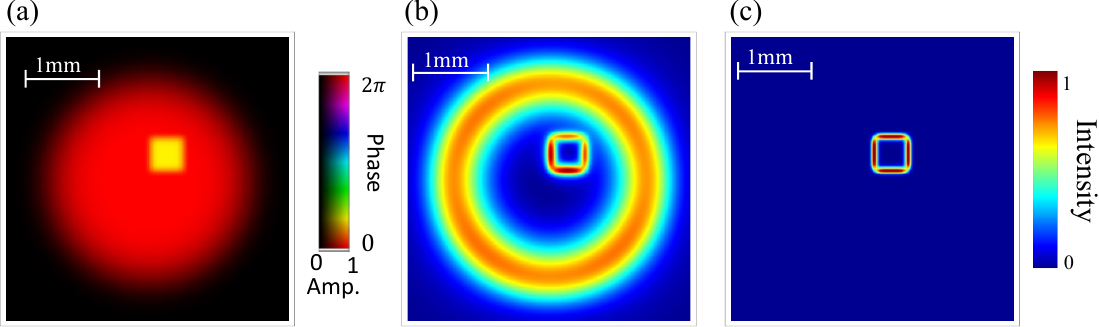}
\caption{(a) Complex amplitude distribution of PAO, (b) its edge-enhanced distribution with on-axis q-plate, and (c) ideal PO edge.}
\label{D1_1}
\end{figure}

\begin{figure}[htp!]
\centering
\includegraphics[width=\linewidth]{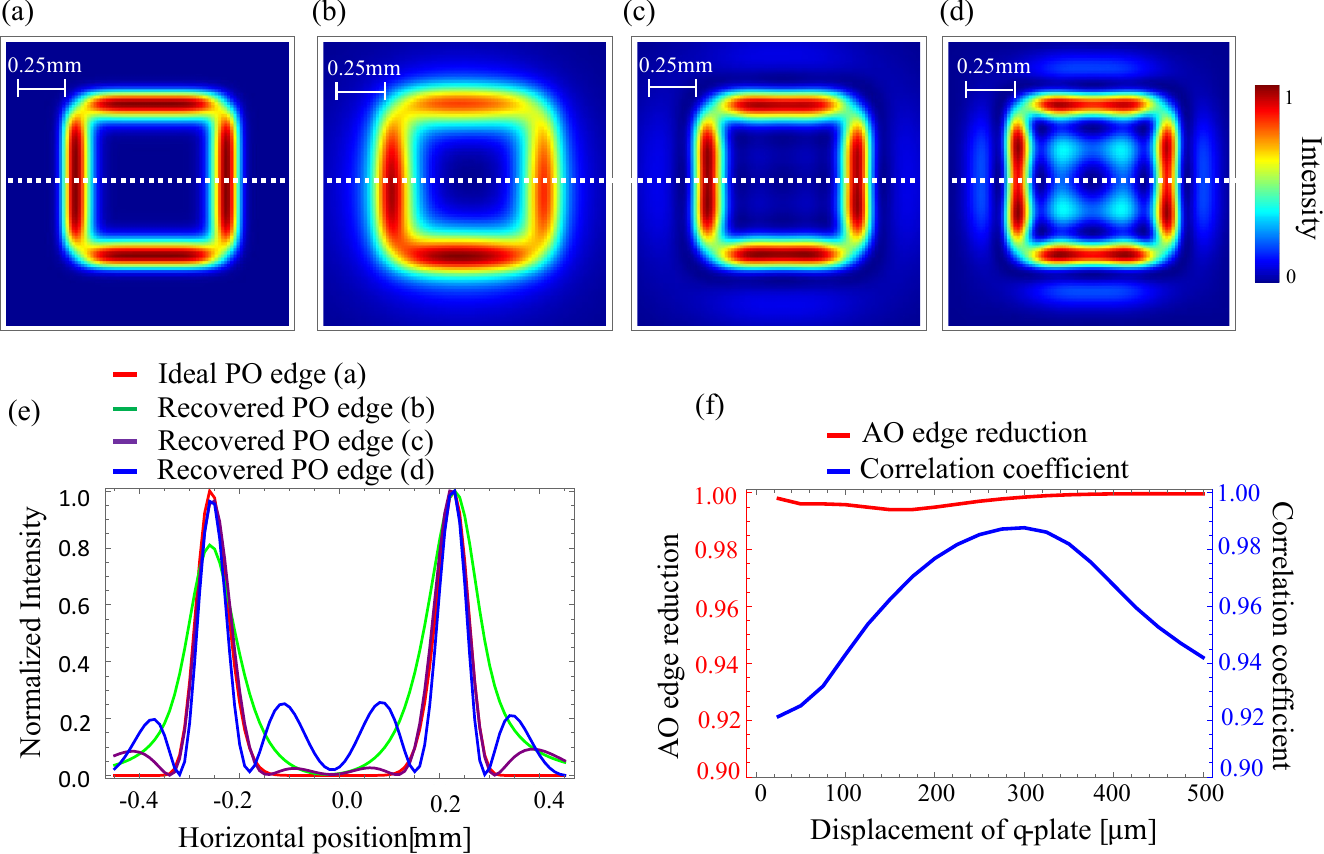}
\caption{(a) The enlarged version of Fig. \ref{D1_1} (c). (b-d) The isolated PO edges using displacement of q-plate by 25, 300, and 500 \textmu m. (e) The horizontal cross-sectional intensity distribution along the ideal PO edge of Fig. \ref{D1_2} (a) corresponds to the red curve, while the green, purple, and blue curves correspond to the isolated PO edges for Figs. \ref{D1_2} (b), (c), and (d), respectively. (f) AO edge reduction in the off-axis q-plate image compared with the on-axis q-plate image and the correlation coefficient between the ideal PO image (Fig. \ref{D1_2} (a)) and PO image isolated by the off-axis q-plate for different q-plate displacements.}
\label{D1_2}
\end{figure}
Now we consider to optimize displacement of an off-axis q-plate in the second $4f$ system to obtain the highest correlation coefficient between the isolated PO and ideal PO edges. We assumed a similar PAO target as mentioned in Section \ref{POaperture}, i.e., the radius of the circular aperture (AO) and the length of the square shape in the phase test target are 1.45 mm and 300 \textmu m through magnification by the first $4f$ system, respectively, and their edges are modeled by using super Gaussian function of order 3. Then, they were multiplied by shifting the square PO by 0.25 mm in the $x$ direction and 0.375 mm in the $y$ direction from the center (see Fig. \ref{D1_1} (a)). The simulation dimension and pixel size were $5 \times5$ mm and $0.01 \times0.01$ mm, respectively. We used a q-plate in the Fourier domain to prepare simulation similar to our experimental setup. We Fourier-transformed the object, multiplied it with the q-plate function, and inversely Fourier-transformed the modified spectrum to achieve the on-axis edge enhancement of the object. We used the ideal q-plate function with omitting $1/r$ convolution to achieve the ideal edge enhancement of the PO. To isolate the PO edge from the AO edge, we used $k_0=\frac{k}{f_1}r_0$ to shift the q-plate, where $f_1$ denotes the focal length of Lens L1 and $r_0$ denotes the shift in the spatial domain. Figures \ref{D1_1} (b) and (c) show the edge enhancement of the PAO using an on-axis q-plate including $1/r$ convolution term and the ideal PO edge enhancement with omitting $1/r$ convolution, respectively.

Figure \ref{D1_2} (a) shows the enlarged version of Fig. \ref{D1_1} (c), which is the ideal PO edge enhancement with omitting $1/r$ convolution. Figures \ref{D1_2} (b), (c), and (d) show the isolated PO edges using displacements of q-plate by 25, 300, and 500 \textmu m, respectively. In Fig. \ref{D1_2} (e), the red curve corresponds to the horizontal cross-sectional intensity distribution of the ideal PO edge in Fig. \ref{D1_2} (a), while the green, purple, and blue curves correspond to the isolated PO edges in Figs. \ref{D1_2} (b), (c), and (d), respectively. Figure \ref{D1_2} (f) shows the AO edge reduction in the isolated PO image and the correlation coefficient between the ideal PO (Fig. \ref{D1_2} (a)) and the isolated PO edges for different displacement of the q-plate. The AO edge reduction remains almost constant for different q-plate displacements over 0.99. However, the correlation coefficient increases at a displacement ranging from 25 \textmu m to 300 \textmu m and then decreases as the displacement of the q-plate increases. 

Since there is no convolution with $1/r$ in the ideal PO edges (Fig. \ref{D1_2} (a)), the red curve in Fig. \ref{D1_2} (e) has the thinnest width of the phase edges. On the other hand, the other three cases are affected by the convolution with $1/r$, while their broadening effect in the width can be reduced by setting the appropriate q-plate displacement $r_0$. The purple curve in Fig. \ref{D1_2} (e) with the highest correlation coefficient (0.987) has a more uniform and thinner phase edge width than the green curve with a smaller q-plate displacement and a lower correlation coefficient (0.921). For further increased q-plate displacement, the phase edge becomes thinner as shown by the blue curve, whereas undesired side lobes surround the phase edges as shown in Fig. \ref{D1_2} (d) that are not present in the other curves. The presence of side lobes in Fig. \ref{D1_2} (d) decreases the correlation coefficient (0.942) between Fig. \ref{D1_2} (d) and the ideal PO edge in Fig. \ref{D1_2} (a). When the q-plate moves away from the beam radius on the Fourier plane, the edge-enhanced microscope is unable to detect the edge of the object. The numerical simulation results indicate that the optimal displacement of the q-plate, defined by a correlation coefficient above 0.98 and edge reduction exceeding 0.98 lies between 250 and 350 \textmu m.

\subsection{ Recovering the PO edge overlapping with the AO}
\label{D2}
   
\begin{figure}[htp!]
\centering
\includegraphics[width=\linewidth]{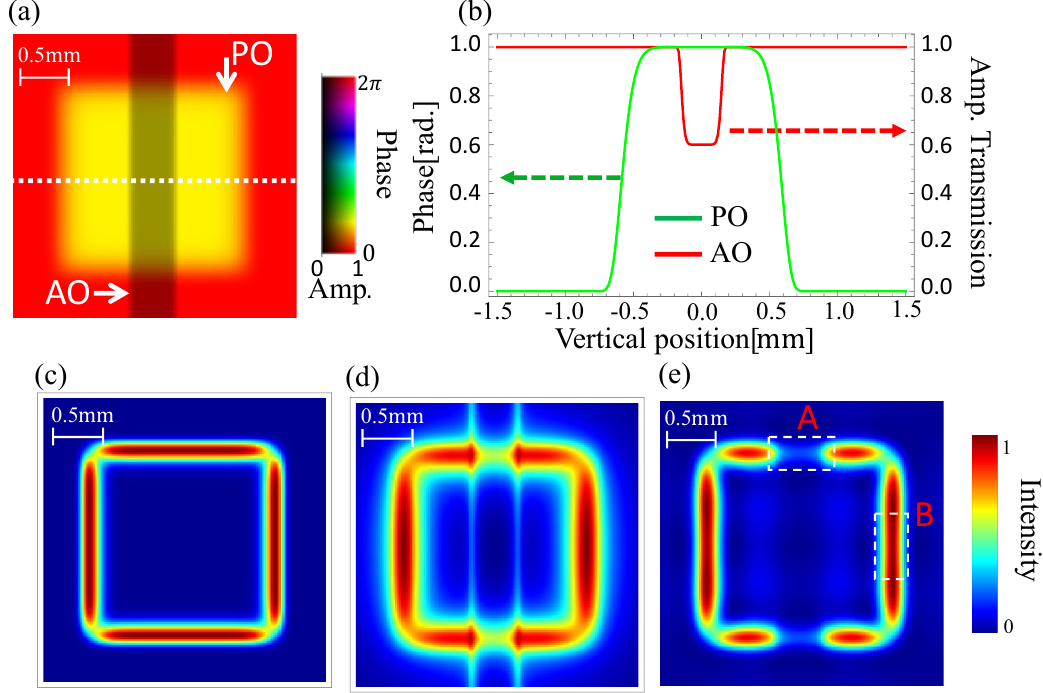}
\caption{(a) Distribution of the PAO, (b) horizontal cross-sectional intensity distributions of the PAO, (c) ideal edge enhancement of the PO, (d) edge enhancement of the PAO using on-axis q-plate, and (e) the isolated PO edge.}
\label{D2_1}
\end{figure}

According to Eq. (\ref{Eq6}), the intensity of the isolated PO edge is proportional to $A(\bm{r})^4$. If the transmittance of the AO is sufficiently high, the PO edge can be recovered by performing division by $A(\bm{r})^4$. However, the lower the transmittance, the more difficult it becomes to recover the PO edge due to noise. Furthermore, the convolution with $1/r$, omitted in Eq. (\ref{Eq6}), makes the PO edge recovery by simple division more difficult.

To investigate the effect of AO transmittance and the convolution with $1/r$ on PO edge isolation behavior, we conducted numerical simulations using a super Gaussian function of order 5 to create a square phase object and a white-hair-like model as shown in Fig. \ref{D2_1}(a). Figure \ref{D2_1}(b) shows the horizontal cross-sectional profile of Fig. \ref{D2_1}(a), with an AO intensity transmittance set to 60\% (as determined from Fig. \ref{fig5}(d)) and a PO phase distribution changing from 0 to 1 radian. Figures \ref{D2_1} (c) and (d) show the ideal PO edge enhancement and the PAO edge enhancement using an on-axis q-plate, respectively. Figure \ref{D2_1} (e) shows the isolated PO edge using off-axis q-plates, which is not fully recovered at point A owing to its overlap with AO. To quantitatively evaluate the accuracy of the recovered PO edge, we use correlation coefficient and ratio between averaged intensities within PO edge areas A and B, as shown in white dot rectangles in Fig. \ref{D2_1} (e). The correlation coefficient between Figs. \ref{D2_1} (c) and (e) is 0.88 and the PO edge ratio A/B in Fig. \ref{D2_1} (e) is 0.13. 

To reduce the broadening effect by $1/r$ convolution in Eq. (\ref{abs}), we adopted the inverse filtering deconvolution as described below. The convolution operation between signal function $f(\bm{r})$ and filter function $h(\bm{r})$ is given by 
\begin{equation}
  g(\bm{r})=f(\bm{r})\ast h(\bm{r}),
  \label{con}
\end{equation}
where $g(\bm{r})$ is the convolved data. For the known $h(\bm{r})$, the deconvolved function $\hat{f}(\bm{r})$ is given by
\begin{equation}
  \hat{f}(\bm{r})=\mathcal{F}^{-1} \left\{G(\bm{k}){H(\bm{k})}^{-1}\right\},
  \label{decon}
\end{equation}
where $\bm{k}$ denotes the frequency domain counterparts of the space variable $\bm{r}$; $G(\bm{k})$ and $H(\bm{k})$ are the two-dimensional Fourier transform of $g(\bm{r})$ and $h(\bm{r})$. 

\begin{figure}[htp!]
\centering
\includegraphics[width=0.50\linewidth]{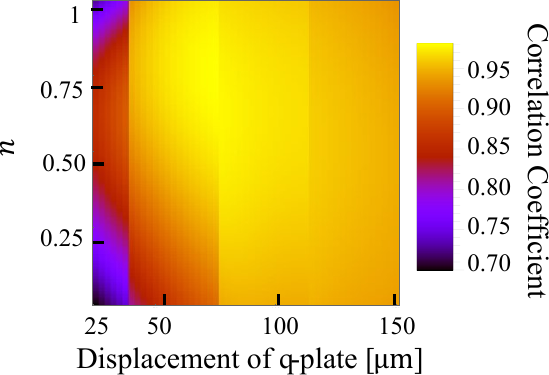}
\caption{The correlation coefficient between the deconvolved PO edge and ideal edge enhancement plotted as displacement of q-plate $k_0$ on the horizontal axis and deconvolution power $n$ in ${\rho}^{n}$ on the vertical axis.}
\label{optimized_k0_n}
\end{figure}

\begin{figure}[htp!]
\centering
\includegraphics[width=\linewidth]{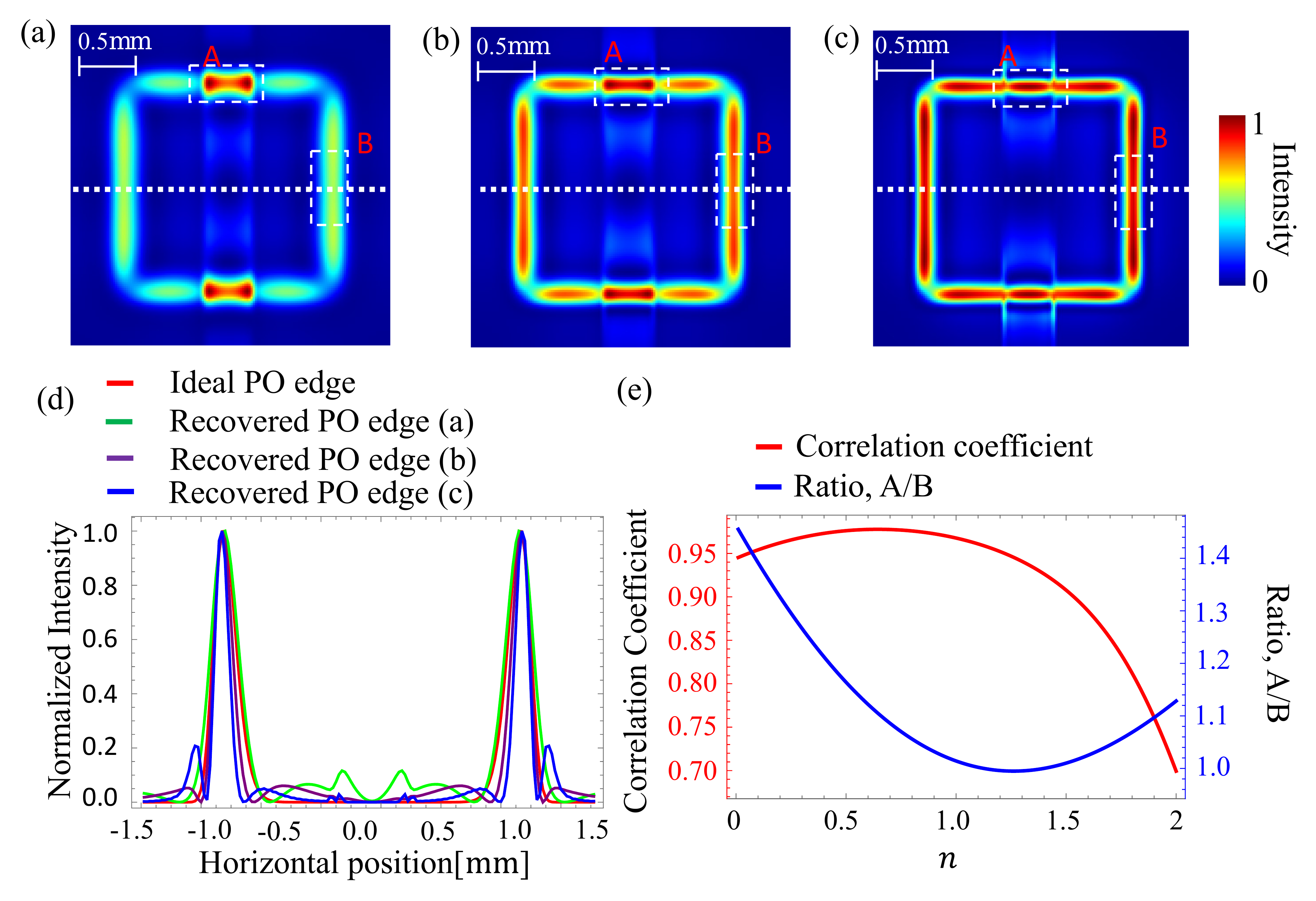}
\caption{(a), (b) and (c) The deconvolved PO edges using $n=0$, 0.64 and 1.12, respectively (displacement of q-plate: 77.5 \textmu m). The deconvolved PO edge with $n=0$ corresponds to the isolated PO edge divided by $A(\bm{r})^4$ without using deconvolution. (d) The horizontal cross-sectional intensity distribution along the ideal PO edge of Fig. \ref{D2_1} (c) corresponds to the red curve, while the green, purple, and blue curves correspond to the deconvolved PO edges for Figs. \ref{optimized-d} (a), (b), and (c), respectively. (e) Correlation coefficient between ideal PO edges and deconvolved PO edges for different $n$ in $\rho^n$ and PO edge ratio A/B in deconvolved PO edges for varying $n$.}
\label{optimized-d}
\end{figure}

Unfortunately, in the edge-enhanced microscope, simple deconvolution technique can not completely recover the signal because the convolution with $1/r$ is applied to the complex amplitude distribution, not the obtained intensity distribution on the camera, as shown in Eq. (\ref{abs}). However, since the convolution with $1/r$ corresponds to $1/\rho$ multiplication in the Fourier domain with $\rho=\sqrt{\smash{k_x}^2+\smash{k_y}^2}$, it is expected that the convolution with $1/r$ can be counteracted by reducing lower frequency and amplifying higher frequency components. Thus, we performs an $\rho^n$ multiplication in the Fourier domain, where $n$ is a parameter to control the extent to which lower frequencies are reduced and higher frequencies are amplified, thereby improving the deconvolved result.

We applied inverse filtering deconvolution technique to output intensity of the edge-enhanced microscope using different values of $n$ in $\rho^n$. We performed a Fourier transform on Eq. (\ref{abs}), multiplied it by $\rho^n$, and finally applied inverse Fourier transform to obtain the deconvolved data $\hat{I}_\text{out} (\bm{r})$ as
\begin{equation}
  \hat{I}_\text{out} (\bm{r})
    \approx \mathcal{F}^{-1} \left\{ \rho^n \mathcal{F}\left\{I(\bm{r})\right\}\right\}.
 \label{Eq11}
\end{equation}
The above deconvolution process was applied to all $I_\text{out} (\bm{r},k_0)$ in Eq. (\ref{Eq6}) and the resultant $J(\bm{r},k_0)$ with the deconvolved intensities is divided by $A(\bm{r})^4$ to recover the PO edge overlapped with the AO.

We optimized the displacement of q-plate $k_0$ and the deconvolution power $n$ in $\rho^n$. Figure \ref{optimized_k0_n} show the correlation coefficient between the recovered PO edge and ideal edge enhancement for different displacement of q-plate $k_0$ and deconvolution power $n$ in $\rho^n$. The displacement of q-plate at 75 \textmu m offers the highest correlation coefficient 0.98 whereas the best PO edge ratio A/B with the displacement of 75 \textmu m was limited to 1.228, which is higher than the ideal value 1. Thus, in what follows, we use the second best displacement 77.5 \textmu m, which offers correlation coefficient 0.97 and the best PO edge ration between A/B reaches almost 1.

%but the ratio between area B and A is 1.3 which higher than 1. Therefore, we consider the trade-off between correlation coefficient and ratio. The ratio between area B and A must be close to 1 for the recovered PO edge. We selected the second best displacement 77.5 \textmu m which offer correlation coefficient 0.97 and ratio between area B and A is 1.1 which is closer to 1. We further optimized deconvolution power $n$ in $k_r^n$ (from 0.01 to 2) using the two best q-plate displacement 75 and 77.5 \textmu m. For the q-plate displacement 75 \textmu m, the ratio between area B and A remain 1.2 whereas for q-plate displacement 75 \textmu m, the ratio between area B and A reaches to 1.0005. Therefore, we consider q-plate displacement 77.5 \textmu m to recovered PO edge at the overlap position with an AO.

Figure \ref{optimized-d} (a) shows the isolated PO edge divided by $A(\bm{r})^4$ without using deconvolution, which corresponds to $n=0$, and Figs. \ref{optimized-d} (b) and (c) show the deconvolved PO edge with $n=$ 0.64, and 1.12, which correspond to the highest correlation coefficient 0.977 and the best PO edge ratio A/B 1.0005, respectively. In Fig. \ref{optimized-d} (d), the red curve corresponds to horizontal cross-sectional intensity distribution of the ideal PO edge in Fig. \ref{D2_1} (c), while the green, purple, and blue curves correspond to the deconvolved PO edges for Figs. \ref{optimized-d} (a), (b), and (c), respectively. For the ideal PO edges, there are no side lobes, unlike the deconvolved PO edges. Figure \ref{optimized-d} (e) shows the PO edge ratio A/B and the correlation coefficient between the deconvolved PO edge and the ideal PO edge in Fig. \ref{D2_1} (c) for the different value of $n$. As we increased the value of $n$ from 0.01 to 2, the correlation coefficient improved, reaching its highest point at $n=0.64$ with a value of 0.977. However, the PO edge ratio A/B is 1.1, which is higher than 1. The PO edge ratio becomes the best ratio of 1.0005 at $n=1.12$ and correlation coefficient is 0.959. These results indicates the deconvolution technique can improve the correlation coefficient and the PO edge ratio A/B.

\section{Conclusion}
Herein, we successfully isolated the PO edge from an AO using off-axis q-plates with four different shifts. The experimental results demonstrated the validity of the proposed theory and method. We quantitatively evaluated the PO edge isolation and AO edge reduction by comparing the results with those obtained from on-axis q-plate images. Moreover, the experimental results were validated via numerical simulations conducted under conditions similar to the experimental setup. The proposed PO edge isolation technique holds great potential for applications in microscopy research and biological edge detection.
Compared to the Transport of Intensity Equation (TIE), which is a commonly used non-interferometric technique for phase retrieval in microscopy research and biological imaging\cite{TIE1defocus,TIE2tilted,TIE4Hilbert,TIE5,TIE3singleshot}, the proposed method offers several advantages. While the TIE techniques require several observations along the propagation direction, the proposed technique only requires four observations and can be performed simultaneously in real time by using a few polarizing beam splitters, waveplates and cameras. Thus, the proposed method is more efficient and practical for real time edge detection in applications. As a solution for the remaining issue of our method with the convolution term $1/r$ in the absolute squared output mentioned in Eq. (\ref{Eq5}), we proposed the inverse filtering deconvolution method and showed its validity from several simulation results. Unfortunately, the noise in actual experimental results that was not considered in our simulations would make it difficult for the proposed method to work well. This issue would be solved with more sophisticated method such as deep learning in the future. Additionally, we plan to investigate in detail the mathematical impact of the $1/r$ convolution term on edge enhancement and phase edge isolation as part of our future work.

\begin{backmatter}
\bmsection{Funding}
Japan Society for the Promotion of Science (18KK0079, 20K05364); Research Foundation for Opto-Science
and Technology.

\bmsection{Disclosures}
The authors declare no conflicts of interest.

\bmsection{Data availability}
Data underlying the results presented in this paper are not publicly available at this time but may be obtained from the authors upon reasonable request.
\end{backmatter}

\bibliography{references.bib}
\end{document}